# Merging droplets in double nano-contact spin torque oscillators

Dun Xiao,[1] Yaowen Liu,[1,*] Y. Zhou,[2,3,*] S. M. Mohseni,[4,5,6] S. Chung,[6,7], and J. Åkerman[6,7,*]

[1.] *Shanghai Key Laboratory of Special Artificial Microstructure Materials and Technology, School of Physical Science and Engineering, Tongji University, Shanghai 200092, China*
[2.] *School of Electronics Science and Engineering, Nanjing University, Nanjing 210093, China*
[3.] *Department of Physics, The University of Hong Kong, Hong Kong, China*
[4.] *Department of Physics, Shahid Beheshti University, Tehran 19839, Iran*
[5.] *NanOsc AB, Electrum 205, 164 40 Kista, Sweden*
[6.] *Materials Physics, School of Information and Communication Technology, KTH Royal Institute of Technology, Electrum 229, 164 40 Kista, Sweden*
[7.] *Department of Physics, University of Gothenburg, 412 96 Gothenburg, Sweden*

## ABSTRACT

We demonstrate how magnetic droplet soliton pairs, nucleated by two separated nano-contact (NC) spin torque oscillators, can merge into a single droplet soliton. A detailed description of the magnetization dynamics of this merger process is obtained by micromagnetic simulations: A droplet pair with a steady-state in-phase spin precession is generated through the spin-transfer torque effect underneath two separate NCs, followed by a gradual expansion of the droplets volume and the out phase of magnetization on the inner side of the two droplets, resulting in the droplets merging into a larger droplet. This merger occurs only when the NC separation is smaller than a critical value. A transient breathing mode is observed before the merged droplet stabilizes into a steady precession state. The precession frequency of the merged droplet is lower than that of the droplet pair, consistent with its larger size. Merged droplets can again break up into droplet pairs at high enough magnetic field with a strong hysteretic response.

---

[*] Author to whom correspondence should be addressed. Email: yaowen@tongji.edu.cn (Y. Liu), yanzhouy@hotmail.com (Y. Zhou), johan.akerman@physics.gu.se (J. Åkerman)

Topological solitons in magnets[1,2] with spin textures have recently attracted considerable attention for spintronic applications.[3-10] Generally, the topological protection makes the lifetime of these nontrivial quasiparticles much longer than that of topologically trivial particles. Magnetic bubbles,[11-13] domain walls,[14,15] magnetic vortices,[16-20] and skyrmions[6-9,21] are typically static topological solitons and have been well studied. Differing from these static structures, the dissipative magnetic droplet soliton (droplet hereafter) is non-topological, inherently dynamic, and results from the balance between anisotropy, exchange, spin transfer torque, and magnetic damping.[22,23]

Droplets have been experimentally reported in nano-contact spin torque oscillators (NC-STOs) having a free layer with strong perpendicular magnetic anisotropy (PMA) and an in-plane spin polarizer fixed layer.[24-27] In such NC-STOs[28-30] the energy dissipated by damping is compensated for by the energy input from the current-induced spin-transfer torque (STT) effect,[31,32] so that the droplet is expected to be strongly localized in the NC region and to have its spins precessing in phase around the film normal with a very large precession angle. The droplet typically has a partially reversed magnetization directly underneath the NC and a zone of large amplitude precession in a region bounding the reversed magnetization, which can lead to an increase in the microwave output power of NC-STOs by a factor[25] of 40 compared to those of non-droplet counterparts. NC-STOs exhibiting droplets hence provide a promising route to achieving much larger signal-to-noise ratios, which is an essential prerequisite for their application.[33] Additionally, the controllability of droplets through current and field may provide a means for the transportation and storage of information with droplets as carriers in future spintronic systems,[23,34] analogous to optical solitons in fiber optic communications.

In this paper, we report on the merging of two individually driven droplets into a single merged droplet by the combined action of an external magnetic field $H$ and a drive current $I$. We find that the merger depends both on the applied current and the magnetic field, as well as on the separation distance between the two magnetic droplets, with the existence of a critical distance $d_c$.

As shown in Fig. 1(a), we consider a NC-STO geometry based on a pseudo-spin-valve structure patterned into a rectangular shape with a long axis of 1024 nm and a short axis of 512 nm. The fixed spin polarizer layer is assumed to be magnetized along the $+z$ direction and the 2-nm thick free layer has perpendicular magnetic anisotropy (PMA). The free layer has two NCs, each with radius $r = 50$ nm and with a separation distance varied from 150 nm to 250 nm. Positive current is defined as the flow of electrons from the free layer to the polarizer layer.

Micromagnetic modeling of the above design was performed using the open-source simulation software MuMax3,[35] based on the Landau-Lifshitz-Gilbert equation including the STT term.[31,36]

$$\frac{d\mathbf{m}}{dt} = -\gamma \mathbf{m} \times \mathbf{H}_{eff} + \alpha \mathbf{m} \times \frac{d\mathbf{m}}{dt} + a_J \mathbf{m} \times (\mathbf{m} \times \mathbf{m}_p) \quad (1)$$

where $\mathbf{m} = \mathbf{M}/M_s$ is the normalized magnetization of the free layer, $M_s$ is the free layer saturation magnetization, $\alpha$ is the Gilbert damping factor, and $\gamma$ is the gyromagnetic ratio. $\mathbf{H}_{eff} = -(1/\mu_0 M_s) \, \partial E/\partial \mathbf{M}$, is the effective magnetic field derived from the free energy density of system. The energy includes the dipole-dipole interaction $E_d$, the exchange interaction parameterized by an exchange constant $A$, the PMA energy, and the Zeeman energy due to an applied magnetic field **H**. The total energy is written as $E=E_d + A[(\partial_x \mathbf{m})^2 + (\partial_y \mathbf{m})^2] - \mathbf{M}\cdot\mathbf{H} - K_u m_z^2$. The last term of Eq. (1) describes the STT effect. The torque factor $a_J(\theta) = \hbar\gamma g(\theta)J/(2|e|M_s t)$, which depends on the current density $J$, the thickness of free layer $t$ and $g(\theta)$. Here $g(\theta)$ is a scalar function depending on the polarization of electrons $P$ and the angle $\theta$ between the magnetization vector of the free layer **m** and that of the polarizer layer $\mathbf{m}_p$. In this study, we assume that the current is injected only into the NC region and that the spin polarization is set to $P = 1$. The following standard parameters are used for the free layer:[37] $M_s$ = 716 kA/m (saturation magnetization), $K_u$ = 447 kJ/m$^3$ (magnetic anisotropy), $A$ = 30 pJ/m (exchange stiffness), and $\alpha$ = 0.01 (Gilbert damping). The free layer is divided into 512×256 unit cells, corresponding to a cell size of 2 nm, which is smaller than the exchange length ($\lambda_{ex} = \sqrt{2A/\mu_0 M_s} = 9.7$ nm). All the simulations are performed at zero temperature.

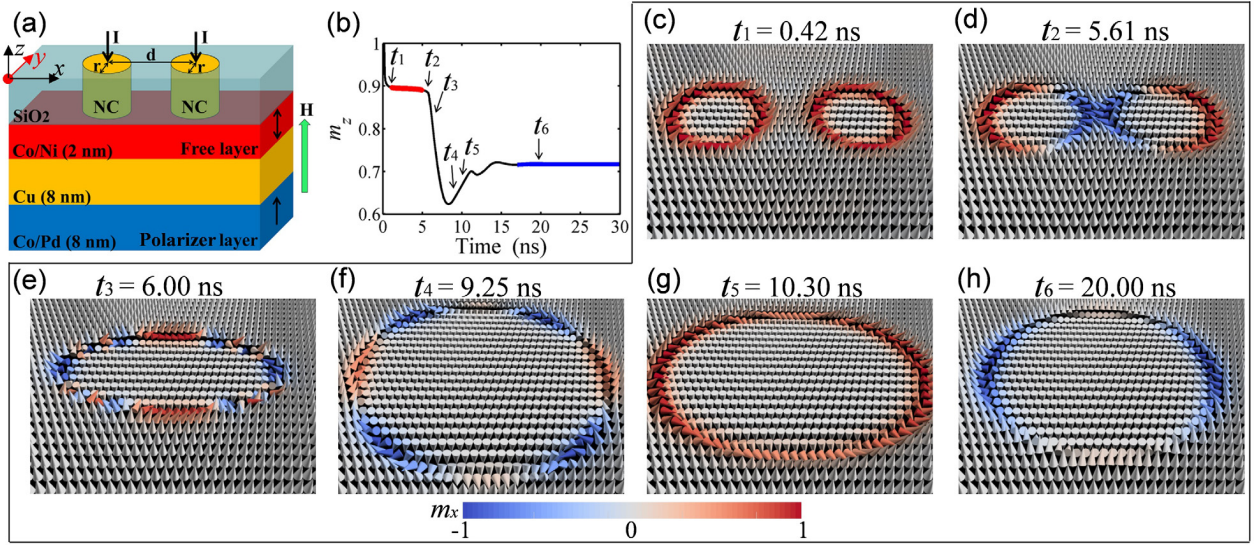

FIG. 1. (Color online) Current- and magnetic field-induced merging of a droplet pair. (a) Schematic illustration of the NC-STO. A magnetic field is applied in the out-of-plane direction with a strength of 0.5 T. The current through each NC is fixed at $I = -15$ mA. (b) Time evolution of $z$-component of average magnetization. (c)-(h) Transient snapshots of magnetization are shown for a small region around NCs. The colored cones display the $x$ component of the magnetization.

A typical example of droplet pair creation and their transformation into a merged droplet is studied in Fig.1(b)-(h), beginning from an initial uniform parallel configuration ($m_z$=+1) with respect to the magnetization direction of the polarizer. The NC separation distance is $d$ = 195 nm, the applied magnetic field is $\mu_0 H$ = 0.5 T (i.e $H$ = 400 kA/m), and current through each NC is -15 mA, resulting in a current density of $1.91\times 10^{12}$ A/m$^2$. Fig. 1(b) shows the time evolution of the $z$-component of the average magnetization over the full free layer area. A series of transient snapshots taken from the magnetization evolution process is shown in Fig. 1(c)-(h), which clearly displays the typical magnetization spatial distribution. The corresponding times are indicated in the temporal magnetization curve of Fig. 1(b). Under the action of the current-induced STT effect, the magnetization at the two NC regions first departs from the initial parallel state and rotates around the $z$-axis. Then, the magnetization at the center region of the NCs is switched to the negative $z$ direction, producing a magnetic droplet pair structure, as shown in Fig. 1(c) at $t_1$. Note that the two droplets have the same precession phase, i.e. they are mutually synchronized.

The two droplets continue to precess as individual droplets for about 5 ns. However, during this time, their radius gradually increases, as evidenced by the gradual reduction in $m_z$ in Fig.1(b) (red line) until their boundaries interact and eventually touch. Owing to the relatively small separation distance, the spin precession underneath the NCs strongly affects the middle region between the droplet pairs through the exchange coupling and the dynamical dipolar interactions. However, unlike the spins underneath the NCs, the spins between the droplets are not directly driven by STT (no current flows in this zone). As a result, the precession phase is delayed to such a degree that the time-dependent magnetization at the inner side (blue color) of the two droplets is opposite in phase to that at the outer side (red color; see Fig. 1(d)). This marks the onset of the two droplets merging to form a single droplet at the moment $t_3$; see Fig. 1(e). The topography of the $m_z$ component during the merging process is shown in the supplementary materials [Fig. S1]. As a result, the reversed magnetization area with $-m_z$ becomes much larger, leading to a sudden drop in the average $m_z$ curve, as shown in Fig. 1(b).

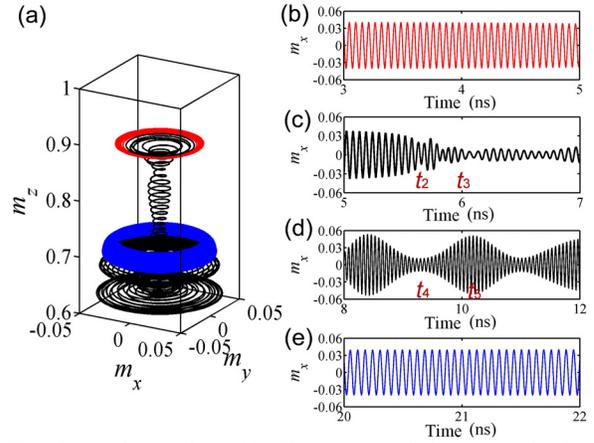

FIG. 2. (Color online) (a) Time trace of the magnetization averaged over the simulation area; the magnetization turn from droplet pair precession orbital (red) to merged droplet precession orbital (blue). (b–e) Magnetization $x$-component evolution for four typical precession states during the merger process.

Interestingly, the precession phase of the spins at the perimeter of the newborn merged droplet exhibits blue-and-red stripes (out of phase; see Figs. 1(e) and (f)), indicating high wave vector perturbations of the perimeter spins. The calculated topological number for this type of merged droplet is however still zero. These strong, high wave vector perturbations are accompanied by strong breathing of the merged droplet. Over time, the blue-and-red stripes, the high wave vector perturbations, and the breathing are all damped out leaving behind a stable droplet with a perimeter having all the spins precessing in phase [Fig. 1(g)].

In order to further investigate the internal mechanism of the merging process, the trace of the magnetization averaged over the full free layer for the same simulation is shown in Fig. 2(a), where the steady-state precession orbit of the droplet pair is indicated by the red curves and that of the merged droplet is indicated by the blue curves. Fig. 2(b)–(e) display the time evolution of the magnetization's $x$-component $m_x$ for four typical precession states, as described in Fig. 1 during the merging process. The perfect stable periodic precession of the droplet pair is given in Fig. 2(b), which corresponds to the state of Fig. 1(c). Fig. 2(c) shows the process when the two droplets merge. The oscillation amplitude of the average $m_x$ decreases slightly when the droplets begin to interact. This is attributed to the fact that the magnetization oscillation in the middle region is opposite in phase to that of the outer side of the two NCs [Fig. 1(c)]. After that, the merged droplet's "breathing mode" state is clearly observable by the average $m_x$, as shown in Fig. 2(d). The moments at which the magnetization configurations of Fig. 1(f) and (g) are taken have been marked in Fig. 2(d). Finally, the merged droplet's breathing mode relaxes and the merged droplet goes into a steady-state periodic precession mode, which can be easily seen in Fig. 2(e).

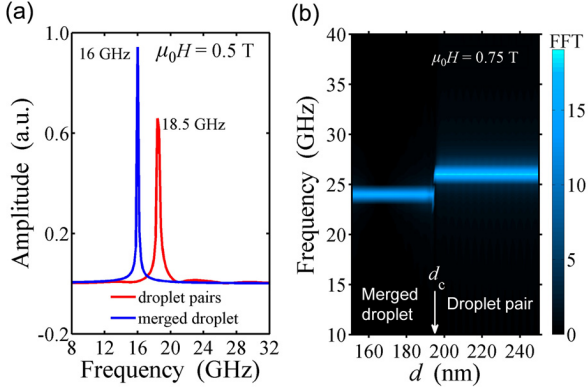
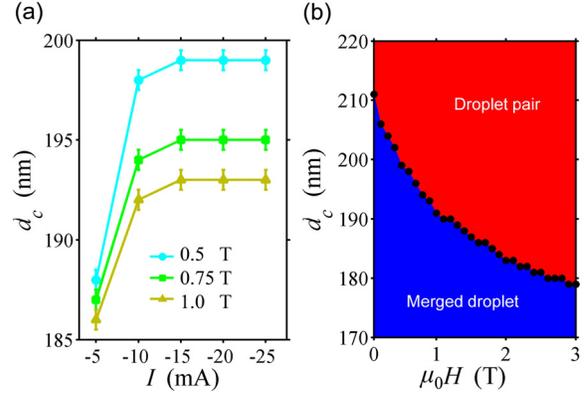

FIG. 3. (Color online) (a) Frequency spectra of the droplet pair and the merged droplet calculated by FFT from the in-plane magnetization $m_x$. (b) Frequency diagram of final precession state for two NCs as a function of separation distance $d$.

FIG. 4. (Color online) (a) The critical separation distance $d_c$ as a function of current for merging a droplet pair into a merged droplet at three different magnetic fields. (b) The operational range as a function of magnetic fields for a given current.

Another interesting result is that the frequency of the merged droplet is a little lower than that of the initial droplet pair, as indicated in Fig. 3(a), showing that the frequency of the merged droplet is 16.0 GHz while that of the original droplet pair is 18.5 GHz. Here the frequencies are calculated from the $m_x$ curves [Fig. 2(b) and 2(e)] using the fast Fourier transform (FFT) method. It should be pointed out that the amplitude of the frequency peaks in Fig. 3(a) calculated by the FFT has no obvious physical meaning. The somewhat larger value of the merged droplet is attributed to its relatively large perimeter, which gives rise to greater $x$-components of the magnetization. It is noticed that the values of the frequency agree well with the relation of solitons:[38]

$$\omega = \omega_0 \sqrt{\pi N_0 / N}. \qquad (2)$$

Here $\omega_0$ is the uniform ferromagnetic resonance frequency. $N = \dfrac{M_s}{2\mu_0}\int(1-m_z)d^3x$ is the number of spin deviations, which can be calculated from the droplet profiles as shown in the supplementary Fig. S1. $N_0$ is the characteristic number of particles. From Fig. 3(a), the precession frequency of the droplet pair $\omega_{pair}$ is 18.5 GHz. For the droplet pair we have $N = 0.7$ and for the merged droplet $N = 0.9$. Thus, the theoretical prediction of the merged droplet frequency can be obtained from Eq. (2), $\omega_{merged} = 0.886 \cdot \omega_{pair} = 16.3$ GHz, which is in good agreement with our simulation value of 16.0 GHz.

It is reasonable to assume that the merging of the droplet pair depends strongly on the separation distance between the two NCs. To understand how the separation influences the droplet merging process, we carried out a series of simulations at $\mu_0 H = 0.75$ T by changing $d$ from 150 to 250 nm. Fig. 3(b) shows the separation distance dependence of the frequency spectra for the steady-state magnetization procession of the final states after 15 ns simulations. A critical value of the separation distance (here $d_c$ = 195 nm) can be clearly observed, below which the droplet pair can merge into a merged droplet due to the strong exchange and dynamical dipolar interaction effects. The oscillation frequency of the merged droplet is 24 GHz, while the magnetic droplet pair oscillates at 26 GHz.

The critical separation $d_c$ plays an important role in the process of merging of the droplet pairs. We find that $d_c$ decreases with applied magnetic field and increases with current, and saturates at a sufficiently large current. Fig. 4(a) shows the current dependence of $d_c$ for various values of the magnetic fields. Obviously, $d_c$ increases with current up to -10 mA, and afterwards maintains an almost constant value of ~192–197 nm for the three different magnetic fields. This result indicates that the further local enhancement of the STT effect through increasing the current in the NC regions cannot increase the critical separation distance of the droplet pair merging. In contrast, decreasing the magnetic field could increase the threshold separation. A more detailed phase diagram of the dependence of $d_c$ on the magnetic field for a droplet pair (red region) and for a merged droplet soliton (blue region) is given in Fig. 4(b), simulated at a fixed current of $I$ = -15 mA. The largest $d_c$ is 211 nm at zero magnetic field. The operational range for merging a droplet pair into a merged droplet is indicated in Fig. 4(b) by the blue color region.

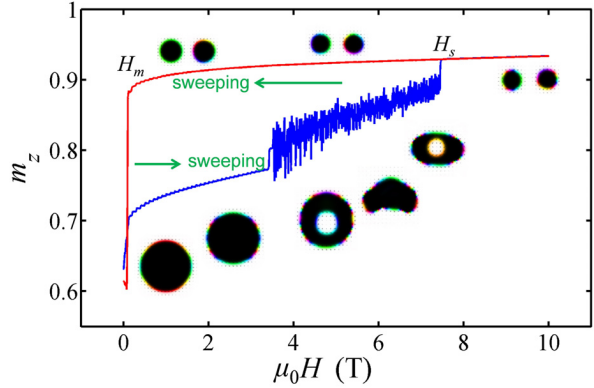

FIG. 5. (Color online) Separation process of a merged droplet into a droplet pair. The blue and red curves show the z-component magnetization as a function of magnetic field sweeping loop. The applied current in each NC is fixed to be -10 mA. The typical magnetization configurations are shown by a series of snapshots during the recover process.

The merger process can also be reversed by increasing the magnetic field to very high values. In other words, the merged droplet can be broken up and the original droplet pair recovered. This recovery process occurs at a comparative high applied field and the droplet pair locally stabilizes in the NCs region. Figure 5 shows a typical recovery process of a droplet pair. The field sweep direction is indicated by the arrows, starting from zero field to 10 T and then back to zero. The blue (red) curve in Fig.5 shows the response of $m_z$ during the $\mu_0 H$ increasing (decreasing) process. The simulation starts from a merged droplet state. A fixed current of -10 mA is applied in each NC. With the field increasing, the spins located in the middle "isolation" zone first reverses to the +$z$ direction, and form a ring-shaped configuration, see Fig.5. But this configuration is extremely unstable and different shaped configurations frequently change when the field increases from 3.5 to 7.5 T. Accordingly, the $m_z$ curve fluctuates widely until the magnetic field larger than a critical magnetic field $\mu_0 H_s$ (=7.5 T), at which the merged droplet is completely separated into a pair of droplets. After that, this droplet pair remains until the magnetic field decreases down to $\mu_0 H_m$ = 0.2 T, at which the droplet pair merges again. We find that the critical field $\mu_0 H_s$ to separate the merged droplet strongly depends on the applied current, showing the decreased $\mu_0 H_s$ for a smaller current (not shown).

Finally, we would like to point out that the initial synchronization state before the merger process is the key factor. Our further simulations indicate that if the two droplets are generated at different moments, the two droplets in some cases will go into a precession state with completely opposite phase. In this case the merger process will not occur. The details of this opposite behavior are however beyond the scope of this work.

In summary, we have investigated the dynamics of a droplet pair merging into a single droplet under specific combinations of applied current and magnetic field. The frequency of the new merged droplet is a little lower than that of the droplet pair in agreement with theory. The merging only occurs for NC separations below a current and field dependent critical value. The inverse process from a larger merged droplet back to a droplet pair can be realized by increasing the external magnetic field. These results provide new insight into a fundamental micromagnetic process that could be useful for spin-torque oscillators and magnonic devices based on magnetic droplet solitons.

This work is supported by the National Basic Research Program of China (2015CB921501) and the National Natural Science Foundation of China (Grant No. 51471118, No.11274241). Y.Z. acknowledges the support by National Natural Science Foundation of China (project No. 1157040329), the Seed Funding Program for Basic Research and Seed Funding Program for Applied Research from the HKU, ITF Tier 3 funding (ITS/171/13, ITS/203/14), the RGC-GRF under Grant HKU 17210014.

**References:**
[1] A. M. Kosevich, B. A. Ivanov, and A. S. Kovalev, Phys. Rep. **194**, 117 (1990).
[2] H.-B. Braun, Advances in Physics **61**, 1 (2012).
[3] V. S. Pribiag, I. N. Krivorotov, G. D. Fuchs, P. M. Braganca, O. Ozatay, J. C. Sankey, D. C. Ralph, and R. A. Buhrman, Nat. Phys. **3**, 498 (2007).
[4] A. Dussaux, et al., Nat. Commun. **1**, 8 (2010).
[5] K. Yamada, S. Kasai, Y. Nakatani, K. Kobayashi, H. Kohno, A. Thiaville, and T. Ono, Nat. Mater. **6**, 270 (2007).
[6] A. Abanov and V. L. Pokrovsky, Phys. Rev. B **58**, R8889 (1998).
[7] B. A. Ivanov, V. A. Stephanovich, and A. A. Zhmudskii, J. Magn. Magn. Mater. **88**, 116 (1990).
[8] A. Fert, V. Cros, and J. Sampaio, Nat. Nanotech. **8**, 152 (2013).
[9] N. Nagaosa and Y. Tokura, Nat. Nanotech. **8**, 899 (2013).
[10] N. L. Schryer and L. R. Walker, J. Appl. Phys. **45**, 5406 (1974).
[11] T. C. Chen and H. Chang, Advances in computer **17**, 223 (1979).
[12] P. Dekker and J. C. Slonczewski, Appl. Phys. Lett. **29**, 753 (1976).
[13] A. P. Malozemoff and J. C. Slonczewski, in *Magnetic Domain Walls in Bubble Materials*, edited by A. P. M.

C. Slonczewski (Academic Press, 1979), p. 217.
[14] A. Thiaville, J. M. García, and J. Miltat, J. Magn. Magn. Mater. **242–245, Part 2**, 1061 (2002).
[15] S. S. P. Parkin, M. Hayashi, and L. Thomas, Science **320**, 190 (2008).
[16] T. Shinjo, T. Okuno, R. Hassdorf, K. Shigeto, and T. Ono, Science **289**, 930 (2000).
[17] K. Y. Guslienko and V. Novosad, J. Appl. Phys. **96**, 4451 (2004).
[18] K. Y. Guslienko and K. L. Metlov, Phys. Rev. B **63**, 100403 (2001).
[19] Y. Liu, S. Gliga, R. Hertel, and C. M. Schneider, Appl. Phys. Lett. **91**, 112501 (2007).
[20] Y. Liu, Z. Hou, S. Gliga, and R. Hertel, Phys. Rev. B **79**, 104435 (2009).
[21] Y. Zhou, E. Iacocca, A. A. Awad, R. K. Dumas, F. C. Zhang, H. B. Braun, and J. Akerman, Nat. Commun. **6**, 8193 (2015).
[22] M. A. Hoefer, T. J. Silva, and M. W. Keller, Phys. Rev. B **82**, 054432 (2010).
[23] M. A. Hoefer, M. Sommacal, and T. J. Silva, Phys. Rev. B **85**, 214433 (2012).
[24] S. M. Mohseni, et al., Science **339**, 1295 (2013).
[25] S. Chung, et al., J. Appl. Phys. **115**, 172612 (2014).
[26] F. Macià, D. Backes, and A. D. Kent, Nat. Nanotech. **9**, 992 (2014).
[27] S. Chung, A. Eklund, E. Iacocca, S. M. Nohseni, S. R. Sani, L. Bookman, M. A. Hoefer, R. K. Dumas, and Å. J., Nat. Commun. (2016) *accepted*.
[28] A. Slavin and V. Tiberkevich, Phys. Rev. Lett. **95**, 237201 (2005).
[29] T. J. Silva and W. H. Rippard, J. Magn. Magn. Mater. **320**, 1260 (2008).
[30] R. K. Dumas, et al., Magnetics, IEEE Transactions on **50**, 1 (2014).
[31] J. C. Slonczewski, J. Magn. Magn. Mater. **159**, L1 (1996).
[32] L. Berger, Phys. Rev. B **54**, 9353 (1996).
[33] S. M. Mohseni, et al., Physica B: Condensed Matter **435**, 84 (2014).
[34] S. Lendínez, N. Statuto, D. Backes, A. D. Kent, and F. Macià, Phys. Rev. B 92, 174426 (2015).
[35] A. Vansteenkiste, J. Leliaert, M. Dvornik, M. Helsen, F. Garcia-Sanchez, and B. Van Waeyenberge, AIP Advances **4**, 107133 (2014).
[36] X. Li, Z. Zhang, Q. Y. Jin, and Y. Liu, New Journal of Physics **11**, 023027 (2009).
[37] E. Iacocca, R. K. Dumas, L. Bookman, M. Mohseni, S. Chung, M. A. Hoefer, and J. Åkerman, Phys. Rev. Lett. **112**, 047201 (2014).
[38] A. M. Kosevich, B. A. Ivanov, and A. S. Kovalev, Physica D: Nonlinear Phenomena **3**, 363 (1981).